\documentclass[prd,twocolumn,a4paper,superscriptaddress]{revtex4}
\usepackage{graphicx}
\begin{document}

\newcommand{\be}{\begin{equation}}
\newcommand{\ee}{\end{equation}}
\newcommand{\bq}{\begin{eqnarray}}
\newcommand{\eq}{\end{eqnarray}}
\newcommand{\bsq}{\begin{subequations}}
\newcommand{\esq}{\end{subequations}}
\newcommand{\bc}{\begin{center}}
\newcommand{\ec}{\end{center}}

\title{String Imprints from a Pre-inflationary Era}

\author{P.P. Avelino}
\email[Electronic address: ]{ppavelin@fc.up.pt}
\affiliation{Centro de F\'\i sica do Porto e Departamento de F\'\i sica da
Faculdade de Ci\^encias da Universidade do
Porto, Rua do Campo Alegre 687, 4169-007, Porto, Portugal}
\affiliation{Astronomy Centre, University of Sussex, Brighton BN1 9QJ, United Kingdom}
\author{C.J.A.P. Martins}
\email[Electronic address: ]{C.J.A.P.Martins@damtp.cam.ac.uk}
\affiliation{Centro de Astrof\'{\i}sica da Universidade do Porto, R. das Estrelas s/n, 4150-762 Porto, Portugal}
\affiliation{Department of Applied Mathematics and Theoretical Physics, Centre for Mathematical Sciences,\\ University of Cambridge, Wilberforce Road, Cambridge CB3 0WA, United Kingdom}
\affiliation{Institut d'Astrophysique de Paris, 98 bis Boulevard Arago, 75014 Paris, France}

\date{12 June 2003}
\begin{abstract}
We derive the equations governing the dynamics of cosmic strings in a flat 
anisotropic universe of Bianchi type I and study the evolution of simple 
cosmic string loop solutions. We show that the anisotropy of the background
can have a characteristic effect in the loop motion. We discuss some 
cosmological consequences of these findings and, by extrapolating our 
results to cosmic string networks, we comment on their ability to survive 
an inflationary epoch, and hence be a possible fossil remnant (still visible 
today) of an anisotropic phase in the very early universe.
\end{abstract}
\preprint{DAMTP-2003-54}
\maketitle

\section{\label{intr}Introduction}

Cosmological inflation \cite{Guth,Albrecht,Linde,Hybrid} 
is a fairly simple paradigm
whose main virtue lies in its ability to solve a number of the problems
of standard cosmology, in particular those related to initial
conditions. Even though it has proven difficult to find a 
single inflationary model that is both (1) well motivated in terms of
a more fundamental theory and (2) in complete agreement with observations,
the inflationary paradigm is broad enough to allow simple toy models
to be in quite good agreement with observational
results \cite{wmap,Peiris}.

Roughly speaking, the way inflation solves these initial condition
problems is by erasing the previously existing conditions and
effectively `re-starting' the universe to a fairly simple state.
Depending on one's point of view, this can be seen either as blessing
or as a curse. The reason for the latter view is that if inflation
is very effective (in practice, if it acts long enough) then
we have essentially no hope of probing the physics of a pre-inflationary
epoch---see \cite{Barrow,Liddle} for enlightening
discussions of these issues.

Fortunately there can be relics left behind after inflation. One possible
example is topological defects \cite{Book}, formed at phase transitions
either before or during inflation \cite{Contaldi,Battye,Openinf}. 
The inflationary epoch will clearly
have the effect of diluting the defect density, and push the network outside
the horizon, freezing it in co-moving coordinates in the
process. However, once the inflationary epoch ends the subsequent
evolution of the defects is necessarily such as to make them come back
inside the horizon \cite{Openinf,Fossils}.

In previous work \cite{Fossils}, we have discussed a specific example
of this behavior. We have considered a domain
wall network produced during an anisotropic phase in the very early universe
(see \cite{Ferreira} for constraints on the present
level of anisotropy),
and shown that in plausible circumstances it could still be present today
and have within it some imprints of the early anisotropic phase.
In the present work, we discuss the analogous scenario for cosmic string
networks. Just as in the domain wall case, we expect such a network to
retain some imprints of such an early anisotropic phase, since it is well
known \cite{Quant,Extending,Moore} that only if it is in a relativistic,
linear scaling regime can such a network erase the traces of its former
conditions. We shall start by discussing cosmic string evolution in a flat,
anisotropic (Bianchi type I) universe in Sect. \ref{csevol}, 
and the evolution of the background in Sect. \ref{bgevol}.
We then study numerically the evolution of cosmic string loops in
Sect. \ref{numsim}, emphasizing the differences with respect to the
standard (isotropic case). Finally in Sect. \ref{net} we comment on
the implications of our results to the evolution of cosmic string networks
as a whole, and we summarize our results in Sect. \ref{conc}. Throughout
the paper we shall work in units where $c=\hbar=1$.

\section{\label{csevol} Cosmic string evolution}

Let us consider the evolution of a cosmic string in a flat anisotropic
universe of Bianchi type I, with line element:
\be \label{3}
ds^2=dt^2-X^2(t)x^2-Y^2(t)dy^2-Z^2(t)dz^2\,.
\ee
Here $X(t)$,$Y(t)$ and $Z(t)$ are the cosmological expansion factors
in the $x$, $y$ and $z$ directions respectively, and $t$ is physical time.
We also define  $A \equiv {\dot X}/X$, $B \equiv {\dot Y} / Y$ 
and $C \equiv {\dot Z}/Z$ where the dot represents a derivative with 
respect to physical time t.

In the limit where the curvature radius of a cosmic string
is much larger than its thickness, we can describe it as a one-dimensional
object so that its world history can be represented by a
two-dimensional surface in space-time (the string world-sheet)
\be\label{1}
x^\nu = x^\nu (\zeta^a)\,; \qquad
a = 0,1\,; \quad\nu = 0,1,2,3
\ee
obeying the usual Goto-Nambu action
\be\label{2}
S=-\mu \int {\sqrt {-\gamma}}d^2 \zeta\,,
\ee
where $\mu$ is the string mass per unit length, $\gamma_{ab}$ is the 
two-dimensional world-sheet metric and $\gamma= {\rm det} (\gamma_{ab})$.
Let us also define
\be\label{def1}
{\bf {\dot x}}^2 \equiv g_{\alpha \beta} 
{\dot x}^{\alpha} {\dot x}^{\beta}=1-X^2 {\dot x}^2-Y^2{\dot y}^2-Z^2 {\dot z}^2\,
\ee
and
\be\label{def2}
{\bf x'}^2 \equiv g_{\alpha \beta} 
x'^{\alpha} x'^{\beta}=-X^2 x'^2 - Y^2 y'^2 - Z^2 z'^2\,,
\ee
so that $\gamma={\bf {\dot x}}^2 {\bf x'}^2$ (using ${\bf {\dot x}} \cdot 
{\bf x'} \equiv g_{\alpha \beta} x'^{\alpha}  {\dot x}^{\beta} = 0$ as a 
gauge condition).

If we choose $\zeta^0=t$ and define $\zeta \equiv \zeta^1$ then the 
string equation of motion is given by \cite{Book}
\begin{widetext}
\be \label{4}
\frac{\partial}{\partial t} \left( \frac{{\dot x}^{\mu} {\bf x'}^2}{\sqrt {- \gamma}}\right)+ \frac{\partial}{\partial \zeta} 
\left( \frac{x'^\mu {\bf {\dot x}}^2}{\sqrt {- \gamma}}\right) +
 \frac{1}{\sqrt {- \gamma}} \Gamma^\mu_{\nu \sigma} ({\bf x'}^2 {\dot x}^{\nu}{\dot x}^{\sigma} + {\bf {\dot x}}^2  x'^\nu  x'^\sigma) = 0\,.
\ee
From the time component we can obtain
\be \label{5}
{\dot \epsilon}+ \epsilon \left[A X^2 \left({\dot x}^2 -  
\frac{x'^2}{\epsilon^2}\right)+B Y^2  \left({\dot y}^2 - \frac{y'^2}{\epsilon^2}\right)+C Z^2  \left({\dot z}^2 - \frac{z'^2}{\epsilon^2}\right) \right]=0\,
\ee
\end{widetext}
where we have made the further definition
\be\label{def3}
\epsilon \equiv {\sqrt {-{\bf x'}^2/{\bf {\dot x}}^2}}={\bf x'}^2
/{\sqrt {-\gamma}= {\sqrt {-\gamma}}/{\bf {\dot x}}^2}\,.
\ee
On the other hand, the $x$ component gives
\be \label{6}
{\ddot x}+ \left(\frac{\dot \epsilon}{\epsilon}+2A\right) {\dot x}+
\frac{1}{\epsilon}\left(\frac{x'}{\epsilon}\right)'=0\,,
\ee
and analogous equations obviously
hold for the $y$ and $z$ components. One can show
that in the limit of an isotropic universe these equations reduce to the
usual form \cite{Book}.

\section{\label{bgevol} Background evolution}

The time component of the Einstein equations in a flat anisotropic universe 
of Bianchi type I is given by
\begin{equation}
{\dot \theta} + A^2 + B^2 + C^2 = - {1 \over 2} k (\rho +3p),
\label{einstein0}
\end{equation}
while the spatial components give
\begin{equation}
{\dot A} + \theta A = {\dot B} + \theta B = {\dot C} + \theta C = 
{1 \over 2} k (\rho -p),
\label{einsteini}
\end{equation}
and we have made the following auxiliary definitions
\be\label{def10}
A=\frac{{\dot X}}{X}\,,\qquad B=\frac{{\dot Y}}{Y}\,, \qquad C=\frac{{\dot Z}}{Z}\,,
\ee
\be\label{def11}
\theta=A+B+C\,,
\ee
and $k=8 \pi G$. It is also useful to combine 
equations (\ref{einstein0}-\ref{einsteini}) to obtain
\begin{equation}
AB+BC+CA=k \rho\,.
\label{einconst}
\end{equation}

In the following discussion we will make the simplification that 
$Z(t)=X(t)$ (and therefore $C=A$) and consider the dynamics of the 
universe during an inflationary phase with $\rho=-p= const$. In this 
case $H^2 \equiv k \rho/3 = const.$ and the Einstein field 
equations (\ref{einstein0}-\ref{einconst}) imply that
\begin{equation}
{\dot A} +{3 \over 2}(A^2 - H^2)=0,
\label{dynA}
\end{equation}
while $B$ can be found from the suggestive relation
\begin{equation}
\frac{B}{A} = \frac{1}{2}\left(\frac{3H^2}{A^2}-1\right).
\label{dynB}
\end{equation}
Equation (\ref{dynA}) has two solutions, depending on the initial conditions.
If $A_i<H$, then $A$ is the smallest of the two dimensions and the shape of
spatial hyper-surfaces is similar to that of a rugby ball. Then the solution
is
\begin{equation}
\frac{A}{H}=\tanh\left[\frac{3}{2}H(t-t_i)+\tanh^{-1}\left(\frac{A_i}{H}\right)\right],
\label{soldynAsmall}
\end{equation}
with $A_i=A(t_i)$.
On the other hand, if $A_i>H$, then $A$ is the larger of the dimensions and
the shape of spatial hyper-surfaces is similar to that of a pumpkin. In that
case the solution is
\begin{equation}
\frac{A}{H}=\coth\left[\frac{3}{2}H(t-t_i)+\coth^{-1}\left(\frac{A_i}{H}\right)\right]\, .
\label{soldynAlarge}
\end{equation}
Note that in both cases the ratio $A/H$ tends to unity exponentially fast, and
hence the same happens with the ratio $B/A$. In other words, inflation tends to
make the universe more isotropic, as expected. An easy way to see
this is to consider the
ratio of the two different dimensions---let us call it $D=B/A$---and
to study its evolution equation. One easily finds
\begin{equation}
{\dot D}=\sqrt{6}H\left(D+\frac{1}{2}\right)^{1/2}(1-D),
\label{evratio}
\end{equation}
which has an obvious attractor at $D=1$.  

Note that even though we have so far assumed (for simplicity)
that $p=-\rho$, the same analysis can be carried out 
for an inflating universe with $p=w \rho$ with $w \neq -1$ by 
numerically solving the conservation equation
\begin{equation}
{\dot \rho} + \theta (\rho+p)=0,
\label{soldynA1}
\end{equation}
together with equations (\ref{einsteini}-\ref{einconst}). Indeed, the
more general case will be relevant for what follows \cite{Fossils}.

\section{\label{numsim}Numerical Simulations}

Let us start by considering the simple case of the evolution of an initially 
static circular cosmic string loop. Its trajectory in the $z=0$ plane can be 
written as
\be
{\bf x}(t,\theta) = q(t, \theta) \left(\sin\theta, \cos\theta, 0\right),
\ee
where $t$ is again physical time. Let us define
\be\label{defrr}
r_x(t,\theta)=X q(t,\theta)\,, \qquad
r_y(t,\theta)=Y q(t,\theta)\,, 
\ee
\be\label{defvv}
v_x(t,\theta)=X \frac{d q(t,\theta)}{dt}\,, \qquad 
v_y(t,\theta)=X \frac{d q(t,\theta)}{dt}\,.
\ee
In the particular case of a spherical loop in a homogeneous and isotropic 
universe $q$ is independent of $\theta$, and hence the evolution equations
become \cite{Quant}
\be
\frac{dv}{dt}=(1-v^2) \left(-\frac{1}{a q}-2Hv\right)\,,
\label{26}
\ee
with $H=A=B=C$, $a=X=Y=Z$ and $v=v_x=v_y$. In what follows we shall numerically
study the evolution of initially static circular loops in a flat 
anisotropic universe, and discuss
the dependence of the results on the background evolution.

\begin{figure}
\includegraphics[width=3.5in,keepaspectratio]{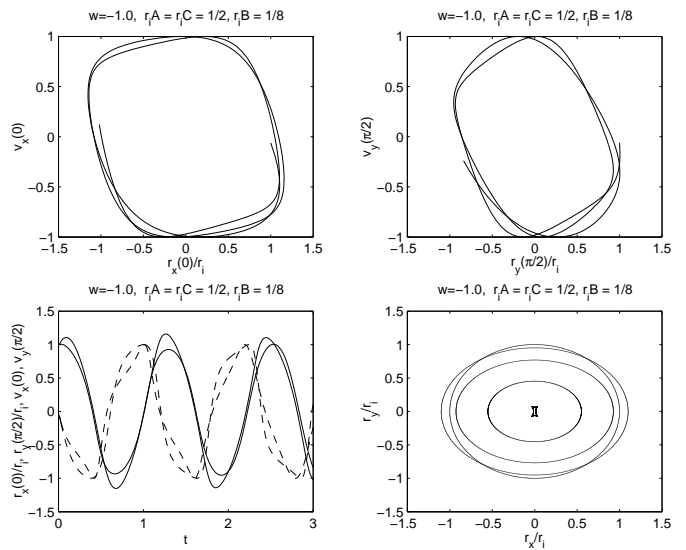}
\caption{\label{figure4}Phase space diagrams (in both the $x$ and the $y$
directions) and time evolution for the loop sizes and velocities, for the
case of a relatively `small' loop. In the bottom left panel
the solid lines represent the loop radii and the dashed lines represent
the loop velocities. See the main text for the definitions
of the various quantities.}
\end{figure}

In Fig. \ref{figure4} we plot the phase space diagrams in both
directions, $(r_x(0)/r_i,v_x(0))$ and 
$(r_y(\pi/2)/r_i,v_y(\pi/2))$, and the corresponding time evolution of the 
loop sizes and velocities ($r_x(0)/r_i$, $r_y(\pi/2)/r_i$, $v_x(0)$ 
and $v_y(\pi/2)$), of an initially static circular loop with 
$r_i=r_x(0)=r_y(0)=0.5/A=0.125/B$ for the $w=-1$ case. We clearly 
see that the motion of the loop in an anisotropic universe is no 
longer periodic, with the anisotropy in the background evolution 
clearly affecting the loop motion. This effect only disappears 
for very small loops ($r_i H \ll 1)$.

\begin{figure}
\includegraphics[width=3.5in,keepaspectratio]{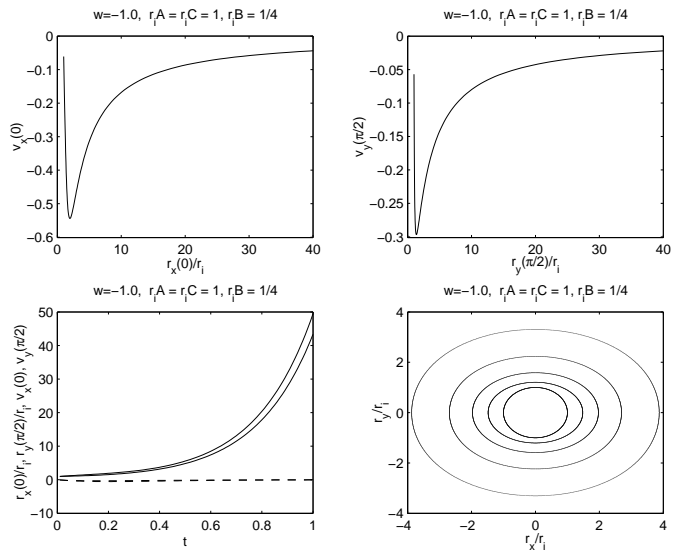}
\caption{\label{figure5}Same as Fig. \protect\ref{figure4}, but for a
larger loop (with twice the size).}
\end{figure}

Fig. \ref{figure5} shows analogous plots but now a larger 
loop (with twice the size) is considered. In this case the loop dynamics 
is dominated by the strong damping caused by the exponential 
expansion, which drives the loop velocity towards zero and freezes the 
loop in comoving coordinates. 

\begin{figure}
\includegraphics[width=3.5in,keepaspectratio]{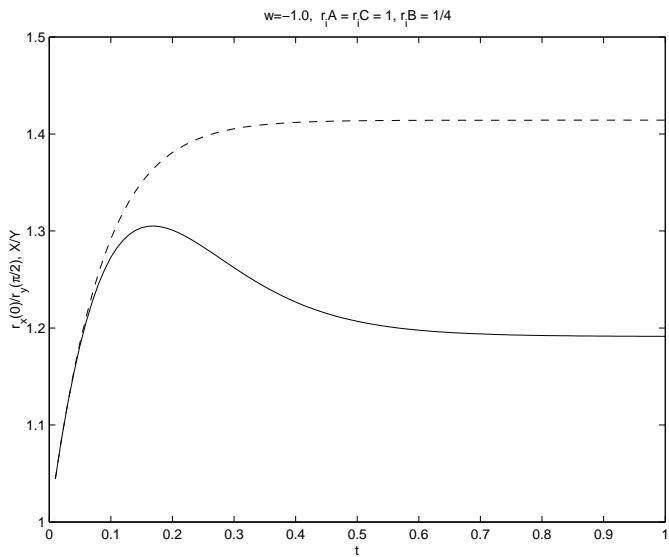}
\caption{\label{figure6}Evolution of the shape of the loop, 
parametrized by $r_x(0)/r_y(\pi/2)$ (solid line), and of the ratio between 
the scale factors in the $x$ and $y$ directions $(X/Y)$ (dashed
line) with physical time, for the larger loop of 
Fig. \protect\ref{figure5}. Here, we assume that $X(0)=Y(0)$.}
\end{figure}

The final shape of loop will be highly asymmetric (see Fig. 
\ref{figure6} for the case of the loop shown in Fig. \ref{figure5}).
The degree of asymmetry will depend both on the initial size of the 
loop and on the asymptotic value of $X/Y$. The asymptotic values 
of the degree of asymmetry (parametrized by $r_x(0)/r_x(\pi/2)$) and 
$X/Y$ will be equal for very large loops. For smaller loops the 
final degree of asymmetry will be smaller, being bound from above 
by $X/Y$. It is also straightforward to show \cite{Fossils} that 
the smaller $w$ is the faster the universe becomes isotropic and the 
smaller is the asymptotic value of $X/Y$. 

Note that although we only study simple loop solutions our main results 
are also expected to hold for the more realistic loops
produced by a cosmic string network.

\section{\label{net}Consequences for Cosmic String Networks}

From the results of the previous section and the basic features of
the standard scenario of cosmic string evolution \cite{Quant,Extending,Moore}
a number of interesting conclusions
can be drawn concerning the evolution of full cosmic string networks in
anisotropic universes.

Firstly our results show that the overall behavior of a cosmic string
network is analogous to that of a domain wall network (discussed in
\cite{Fossils}) in similar circumstances.
The existence of an anisotropic phase through which the
network evolved will be imprinted on it much beyond the time when the
background becomes isotropic. In fact, it will be imprinted on
the network as long as it is frozen
outside the horizon. (Note that in this situation the network will
not seed any density fluctuations---for this to happen other mechanisms
would be required, as discussed in \cite{adiabatic,gaussian}.)
Only when it falls inside the horizon it will start to become
relativistic and isotropic. Again as in the domain wall case, we expect
that the evolution towards the relativistic regime will be somewhat
slower than in the standard case, which could conceivably have
observational implications.

Using the results of \cite{Openinf,Fossils} it is possible to see that a 
cosmic string network can `survive' up to about 60 e-foldings of inflation 
(the exact number being model-dependent), in the sense that any
network produced in such a period will still come back inside the
horizon in time to have observable consequences by the present day.

Of course, when the network starts becoming isotropic after coming 
back inside the horizon the anisotropic signatures will gradually 
disappear, but we still 
expect it to leave an observational imprint of its former state. While it is 
beyond the scope of 
the present article to carry out a detailed analysis of observational 
constraints on this scenario we will nevertheless 
provide a simple discussion of some qualitative features.

The first point to notice is that cosmic strings are much more benign
than domain walls, and hence we expect that constraints on this scenario
(which are fairly tight in the domain wall case) will be very much weaker 
in the context of cosmic strings. There is another major difference 
with respect to the domain wall case. It is well known \cite{Press} that 
domain wall `balls' have a negligible dynamical effect in the evolution of the
network, because they are relatively few and unwind very quickly. In 
contrast, cosmic string loops generally contribute a significant fraction of 
the overall energy density of the string network \cite{Quant} and also may 
have a crucial role in seeding density perturbations \cite{loops1,loops2}. 

Imprints of an anisotropic defect network could conceivably be seen in the 
cosmic microwave background or, at lower redshifts, in a search involving 
gravitational lensing due to cosmic strings---for which there has been a 
recent claim of a detection \cite{lensing}. Clearly further study is required 
if one is to make a quantitative assessment of their observational 
implications.

\section{\label{conc}Conclusions}

We have studied the dynamics and cosmological consequences
of cosmic strings in a flat 
anisotropic universe of Bianchi type I. Focusing on the evolution of
simple loop solutions, we have demonstrated that the anisotropy of the
background has a characteristic effect in the loop motion, in particular
preventing the existence of periodic solutions.

We have also discussed some cosmological
consequences of these findings both for long strings and loops in a
realistic cosmic string network. Much like in the domain wall case
\cite{Fossils}, we have seen that cosmic string networks can 
remain anisotropic much beyond the
epoch when the background becomes isotropic.
The inflationary epoch will push the network outside
the horizon, freezing it in co-moving coordinates and hence freezing the
anisotropy with it. Once the inflationary epoch ends the subsequent
evolution of the defect network  is necessarily such as to make it
come back inside the horizon, but it can only start loosing the anisotropic 
signature once it is unfrozen, \textit{id est} once it's fully inside the 
horizon and relativistic.

Just as in the case of domain walls \cite{Fossils} there are
a number of possible observational signatures of the existence of
such a phase in the long string network itself. 
The detailed study of the observational consequences of this scenario
is beyond the scope of the present work, but clearly deserves further
scrutiny. Finally, let us conclude by
emphasizing that the results we presented are further evidence of
the fact that the importance of cosmic string (and topological defects in
general) as probes of
the physics of the early universe goes well beyond their possible
role in seeding structure formation.

\begin{acknowledgments}
C.M. is funded by FCT (Portugal), under grant FMRH/BPD/1600/2000.
This work was done in the context of the ESF COSLAB network.
Additional support came from FCT under contract CERN/POCTI/49507/2002.
\end{acknowledgments}
\bibliography{imprints}
\end{document}